\documentclass[9pt,twocolumn,twoside]{osa-article}

\journal{boe} 

\usepackage[utf8]{inputenc}
\usepackage[T1]{fontenc}
\usepackage{gensymb}
\usepackage{amsmath}
\usepackage{subfiles}
\usepackage{acro}

\newcommand{\norme}[1]{\left\Vert #1\right\Vert}

\DeclareAcronym{cslo}{
  short = cSLO,
  long  = confocal scanning laser ophthalmoscopy,
  class = abbrev
}
\DeclareAcronym{ism}{
  short = ISM,
  long  = image scanning microscopy,
  class = abbrev
}
\DeclareAcronym{sim}{
  short = SIM,
  long  = structured illumination microscopy,
  class = abbrev
}
\DeclareAcronym{snr}{
  short = SNR,
  long  = signal-to-noise ratio,
  class = abbrev
}
\DeclareAcronym{sio}{
  short = SIO,
  long  = structured illumination ophthalmoscope,
  class = abbrev
}
\DeclareAcronym{amd}{
  short = AMD,
  long  = Age-related Macular Degeneration,
  class = abbrev
}
\DeclareAcronym{dmd}{
  short = DMD,
  long  = digital micromirror device,
  class = abbrev
}

\newcommand{\note}[1]{{\em\textbf{\color{blue}[#1]}}}

\newcommand{\noteylt}[1]{{\em\textbf{\color{red}[#1]}}}
\newcommand{\etal}{{\em{et al.}}}

\renewcommand{\noteylt}[1]{#1} 
\renewcommand{\note}[1]{} 

\newif\ifVersionFinaleNature
\VersionFinaleNaturetrue

\begin{document}

\title{Super-resolution \textit{in vivo} retinal imaging using structured illumination ophthalmoscopy}

\author{Yann Lai-Tim,\authormark{1,2,5,*} Laurent M. Mugnier,\authormark{1,5,*} Léa Krafft, \authormark{1,5} Antoine Chen, \authormark{1,3,5} Cyril Petit, \authormark{1,5} Pedro Mec\^e, \authormark{4,5} Kate Grieve, \authormark{2,5} Michel Paques, \authormark{2,5} and Serge Meimon\authormark{1,5}}

\address{\authormark{1}DOTA, ONERA, Université Paris Saclay, F-92322 Châtillon, France\\
\authormark{2}CIC 503, INSERM, Quinze-Vingts National Ophthalmology Hospital, 28 rue de Charenton, 75012 Paris, France\\
\authormark{3}Quantel Medical, 63808 Cournon d'Auvergne, France\\
\authormark{4}Institut Langevin, ESPCI Paris, CNRS, PSL University, 1 rue Jussieu, 75005 Paris, France\\
\authormark{5}Paris Eye Imaging group, Quinze-Vingts National Ophthalmology Hospital, 28 rue de Charenton, 75012 Paris, France, www.pariseyeimaging.com}

\email{\authormark{*}yann.lai-tim@onera.fr, laurent.mugnier@onera.fr, serge.meimon@onera.fr} 







\begin{abstract}
Structured illumination microscopy (SIM) is one of the most versatile super-resolution techniques. Yet, its application to live imaging has been so far mainly limited to fluorescent and stationary specimens. Here, we present advancements in SIM to jointly tackle all the challenges of imaging living samples, \textit{i.e.}, obtaining super-resolution over an undistorted wide-field while dealing with sample motion, scattering, sample-induced optical aberrations and low signal-to-noise ratio. By using adaptive optics to compensate for optical aberrations and a reconstruction algorithm tailored for a moving and thick tissue, we successfully applied SIM to in vivo retinal imaging and demonstrated structured illumination ophthalmoscopy for high contrast super-resolution \textit{in vivo} imaging of the human retina.
\end{abstract}



\section{Introduction}

The eye is the only organ that permits direct observation at sub-micron wavelength of the human neurovascular network, which is involved in the disease process for a variety of conditions including diabetes~\cite{mizutani_accelerated_1996}, hypertension~\cite{wolf_quantification_1994} and stroke~\cite{girouard_neurovascular_2006}.  When observed with a sufficient numerical aperture, \textit{in vivo} micron-scale imaging of retinal neurons can be achieved, provided that the optical defects of the eye are corrected.  Adaptive optics (AO) has been used to that aim for more than 20 years~\cite{Liang:97} , allowing diffraction-limited retinal imaging and thus revolutionizing the understanding of the structure and function of the normal visual system~\cite{roorda_arrangement_1999}, as well as retinal disease diagnosis and follow-up.
However, the optical aperture of any ophthalmoscope is eventually limited by the eye itself, via the iris. Even with chemically induced mydriasis, the diffraction limit often hinders the resolution of the smallest retinal cells. Several imaging techniques have been developed in microscopy to go beyond the diffraction limit, but only few of them are compatible with \textit{in vivo} retinal imaging: \ac{cslo}~\cite{roorda_adaptive_2002}, \ac{ism}~\cite{dubose_superresolution_2019-1} and \ac{sim}~\cite{gruppetta_theoretical_2011}. 

As noted in 1984 by Wilson \etal~\cite{wilson_theory_1984}, a resolution gain can be obtained with a \ac{cslo} only if the confocal pinhole diameter is smaller than one Airy disk diameter, at the expense of losing photons. It leads to a trade-off of signal-to-noise ratio~(SNR) for resolution, a fundamental drawback that has led to the invention of more photon-efficient strategies, such as ISM~\cite{roth_optical_2013}.  This technique, also known as optical reassignment, is similar to a scanning laser ophthalmoscope where for each position of the illumination spot a wide-field image is acquired. The correct exploitation of these images leads to a potential factor $\sqrt{2}$ gain in resolution, with no photon loss. However, \ac{ism} and \ac{cslo} are both scanning techniques, prone to distortion due to eye motion during the raster scanning of the field of view. In consequence, scanning systems may resort to non-scanning wide-field images as ground truth to dewarp their images~\cite{salmon_automated_2017}. Therefore, there exists an unmet need for a super-resolved retinal imager that would be both photon-efficient and distortion-free.

\begin{figure*}[!bht]
\centering
\includegraphics[width=0.76\linewidth]{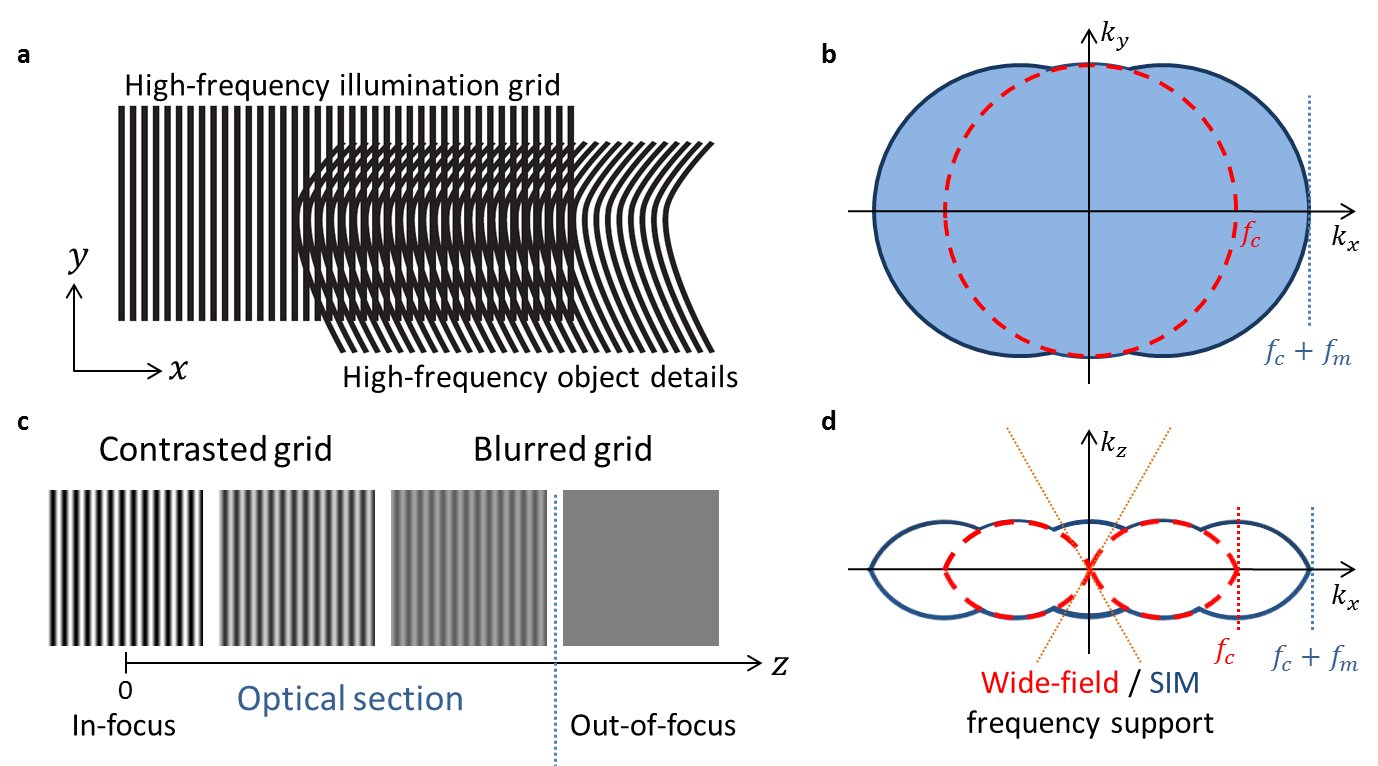}
\caption{\protect\ac{sim} principle.
a, A high-frequency sinusoidal illumination pattern is projected onto the observed object. Due to the Moiré effect (seen here as the apparent horizontal coarser lines in the overlap area), this projection down-modulates high-frequency object details that are optically unseen into the optical accessible bandwidth.
b, This enables the extraction of high-frequency details beyond the classical diffraction limit of optical imaging, which can be seen in Fourier space, as an extended frequency support (blue area) compared with the conventional wide-field bandwidth (red circle). It should be noted that SIM increases the conventional wide-field bandwidth by the value of the grid frequency $f_m$ along its modulation direction. Several orientations of the grid pattern should be used for isotropic super-resolution.
c, Additionally, as the contrast of the illumination pattern in the image focal plane decreases with defocus $z$, it is possible to discriminate the in-focus object content where the illumination grid is high contrast from the out-of-focus background where the grid is not visible due to blurring.  Thus, the out-of-focus background can be computationally removed to reconstruct an optical section of the in-focus content.
d, In the reciprocal space, this optical sectioning capability can be seen in the extended \protect\ac{sim} frequency support, which covers spatial frequencies along the $k_z$ axis that are missing in the conventional wide-field frequency support (yellow cone).}
\label{fig:sim}
\end{figure*}

An alternative to \ac{ism} and \ac{cslo} that achieves both super-resolution and optical sectioning without distortion is SIM~\cite{neil_method_1997,gustafsson_surpassing_2000}. This technique consists of illuminating the observed sample with fringe patterns. By acquiring images for different orientations and positions of the illumination patterns, a super-resolved and optically sectioned image can be reconstructed. Figure~\ref{fig:sim} illustrates the principle of the resolution enhancement and optical sectioning achieved by \ac{sim}.
Recent work~\cite{shroff_structured_2010,gruppetta_theoretical_2011,chetty_structured_2012} has aimed at applying \ac{sim} to human retinal imaging but no experimental validation was provided. We identify three main issues, discussed in the above-mentioned works, which may have so far prevented the successful application of \ac{sim} to \textit{in vivo} retinal imaging.
Firstly, eye-induced optical aberrations hinder the projection of contrasted high spatial frequency illumination patterns onto the retina and reduce the diffraction-limited passband of the instrument~\cite{gofas-salas_high_2018}.
Secondly, the retinal motion stemming from uncontrolled eye movements~\cite{martinez-conde_role_2004} must be taken into account properly to avoid artefacts in reconstructions~\cite{forster_motion_2016}.
Lastly, the retina is a thick and scattering tissue. Consequently, wide-field retinal images have poor SNR due to a strong scattering background. 
Although some of these issues have been addressed in recent work by Turcotte \etal~\cite{turcotte_dynamic_2019} in the case of \textit{in vivo} imaging of mouse brain with submicron residual motion, there is no report of a successful implementation of super-resolution SIM for \textit{in vivo} imaging of thick tissues with significant motion, let alone in \textit{in vivo} human.

In this paper, we describe an implementation of \ac{sim} applied to a living human retina that addresses all of these three issues, enabling super-resolution retinal imaging. 
We have developed a \ac{sio}, combining a custom-made AO system~\cite{gofas-salas_high_2018} and a digital micromirror device (DMD) based illumination to project high spatial frequency fringe patterns onto the retina while mitigating the effects of ocular aberrations as described in the Methods.
Our reconstruction technique takes the object motion and the scattering background into account so as to achieve both super-resolution and optical sectioning on a thick and moving tissue such as the living retina.
Moreover, we propose a SIM acquisition strategy that turns the sample motion from an inconvenience to an asset. 


\section{Methods}

\subsection*{Imaging system}



\begin{figure}[!htb]
\centering
\includegraphics[width=0.9\linewidth]{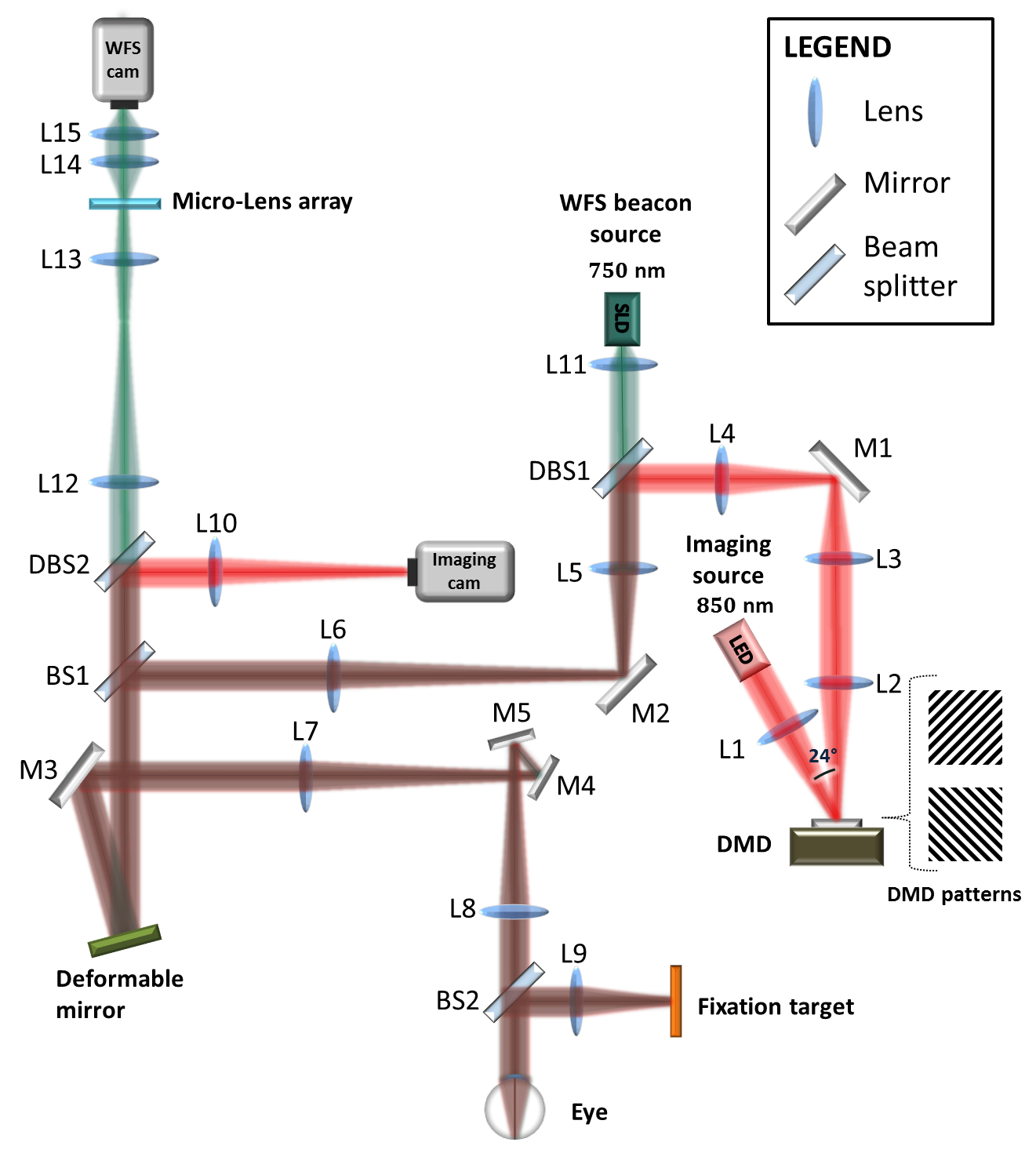}
\caption{SIO schematic. All optical components are labelled: SLD, superluminescent diode; LED, light-emitting diode; DMD, digital micromirror device; L1-L15, lenses; M1-M5, plano-mirrors; BS1-BS2, beamsplitters; DBS1-DBS2, dichroic beamsplitters. The beams illustrated in red, green and brown depict respectively the illumination and detection path, the wavefront sensing path and the common (red + green) path.
}
\label{fig:setup}
\end{figure}

To achieve super-resolution SIM imaging, coherent fringe projection techniques are usually preferred as they yield higher modulation contrast than incoherent projection techniques. However, it was shown that such projection techniques led to high speckle noise when applied to \textit{in vivo} retinal imaging~\cite{shroff_structured_2010}.
Thus, we developed a structured illumination ophthalmoscope~(SIO) using an incoherent projection technique because it produces speckle-free SIM retinal images and it enables better optical sectioning than coherent illumination~\cite{wilson_optical_2011}. The layout of the \ac{sio} is shown in Figure~\ref{fig:setup}.
It is composed of two optical subsystems: the wavefront sensing and control (WFS) subsystem and the imaging subsystem.

The WFS subsystem enables AO correction of the illumination and detection beams and includes a fibered super luminescent diode (SLD) (LEDMOD, Omicron, Germany) centered at 750 nm, a custom-built Hartmann-Shack sensor and a deformable mirror (ALPAO, France). More details about the AO implementation can be found here~\cite{gofas-salas_high_2018}.

The imaging subsystem consists of a DMD-based illumination path that projects fringe patterns onto the retina using an incoherent LED source with a central wavelength of $850$ nm (M850LP1, Thorlabs, USA) and a detection path that directs the light backscattered by the retina toward an imaging camera (ORCA flash4-V2, Hamamatsu, Japan).
In order to produce the illumination patterns, a DMD (V-650L, Vialux, Germany) is used as an amplitude-only spatial light modulator. It consists of an array of $1280\times800$ micromirrors that can be individually tilted along their diagonal to two angle positions: $+12\degree$ tilt reflects the incident beam to the optical axis and $-12\degree$ tilt deflects it away from the optical axis. Since the DMD micromirrors tilt along their diagonal, the DMD chip was rotated $45\degree$ along the optical axis in order to keep both incident and reflected beams in one plane, parallel to the optical table.
The DMD, illuminated by a collimated LED source, was set to reflect a binary fringe patterns of spatial frequency $34$ cycles/degree oriented at $\pm 45\degree$ toward the optical path. These illumination patterns are then demagnified and projected onto the retina through multiple relay optics (lenses L2+L3; L4+L5; L6+L7; L8 + eye lens), fold mirrors (M1-M5) and a deformable mirror, which accounts for ocular aberrations that affect the illumination beam on its way into the eye.
Even though only binary fringe patterns are produced by the DMD, a sinusoidal intensity distribution is obtained in the retina due to the limited optical bandwidth that cuts off the higher order spatial frequencies (harmonics) of the binary patterns.

The imaging camera (ORCA flash4-V2, Hamamatsu, Japan), conjugate with the DMD, collects the light backscattered by the retina enabling $5\degree \times 5\degree$ frame acquisition at 100 Hz. The backscattered beam is corrected for the ocular aberations by the deformable mirror to mitigate their effect on the acquired images. As the projected illumination pattern is focused on the retinal plane conjugate to the DMD, we can image various retinal layers by adding a defocus aberration with the deformable mirror on both the illumination and detection paths using the AO system.

\subsection*{Image acquisition}

Instead of accurately phase shifting the illumination patterns over the observed object as is usually done to obtain the phase diversity required for \ac{sim} reconstruction~\cite{neil_method_1997,gustafsson_surpassing_2000}, the uncontrolled eye movements~\cite{martinez-conde_role_2004} introduce inter-frame retinal shifts that can be exploited to provide this phase diversity~\cite{shroff_lateral_2010, chetty_structured_2012}. The SIO raw images are thus acquired using a static fringe pattern, sequentially oriented at $+45\degree$ and $-45\degree$ to enable two-dimensional super-resolution.
To ensure that the inter-frame retinal shifts provide a sufficient phase diversity~\cite{shroff_lateral_2010,chetty_structured_2012}, at least 6 or 7 frames per orientation of the pattern are captured.

\begin{figure*}[!htbp]
\centering
\includegraphics[width=\linewidth]{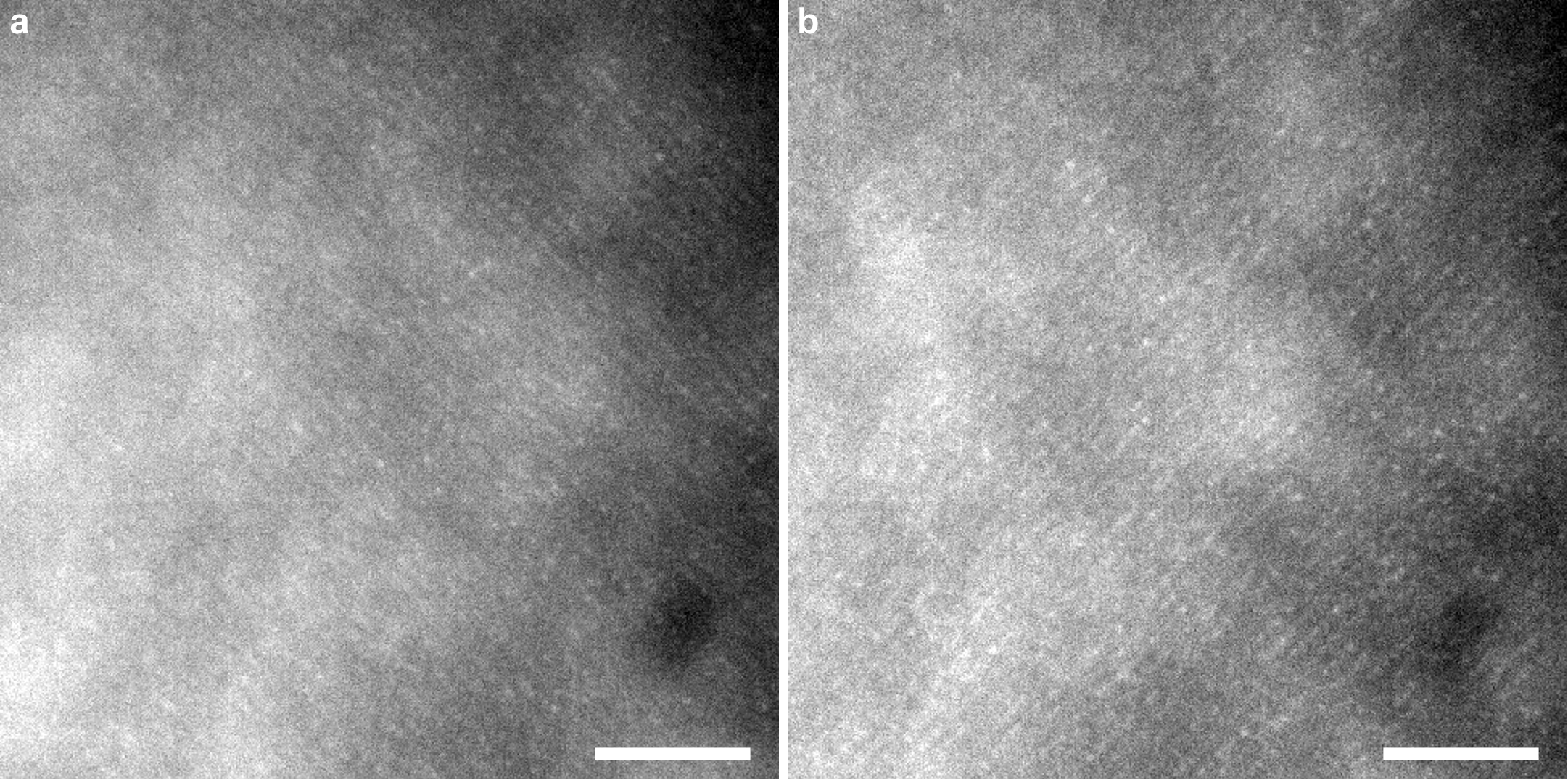}
\caption{$1.24\degree \times1.24\degree$ cropped individual SIO frames. a, SIO frame with a $-45\degree$ oriented fringe pattern, b, SIO frame with a $+45\degree$ oriented fringe pattern. Scale bars, 0.25\degree. The fringe spatial frequency corresponds to $34$ cycles/degree. A linear LUT with a saturation of the 0.3\% of the brightest and darkest pixels was applied to the images.}
\label{fig:sio_frame}
\end{figure*}

Before image acquisition, the onboard memory of the DMD was preloaded with our illumination patterns using the EasyProj application provided by VIALUX. Two fringe patterns were sequentially displayed by the DMD with a switching rate of $20$ Hz. Those patterns were set to project fringes of $34$ cycles/degree spatial frequency in the retina, alternatively oriented at $+45\degree$ and $-45\degree$.
We set the imaging camera to acquire $2048\times2048$ pixel 16-bit images with a frame rate of $100$ Hz and an exposure time of $10$ ms using the HCImage software (Hamamatsu, Japan).
The images were acquired on a healthy subject, who signed an informed consent before participation. The experimental procedures adhered to the tenets of the declaration of Helsinki. The study was authorized by the appropriate ethics review boards (CPP and ANSM (IDRCB number: 2019-A00942-55)).
The subject was seated in front of the SIO and stabilized with a chin and forehead rest. A fixation target was used to guide the subject's line of sight.
The image acquisition was performed in standard conditions with neither pupil dilation nor cyclopegia, in a dark room. In these conditions, the imaging pupil size was limited by the deformable mirror to 7.5 mm diameter in the eye pupil plane, leading to a theoretical diffraction-limited cutoff frequency of $f_{c,\text{diff}}=D/\lambda=154$~cycles/degree, with $D$ the pupil diameter at the eye and $\lambda$ the imaging wavelength.
The optical power at the entrance of the eye from the illumination source and the WFS source is, respectively, $850\mu$W and $1.8\mu$W, which are below the ocular safety limits established by the ISO standards for group 1 devices. Figure~\ref{fig:sio_frame} shows examples of raw SIO frames acquired for both orientations of the illumination pattern.

\subsection*{SIO reconstruction}

Before computing the SIO reconstruction, the SIO raw frames were pre-processed as follows. An average dark frame was subtracted from the raw SIO images in order to suppress the camera offset. Then the images were cropped to $1536\times1536$ pixels and apodized to avoid ringing artefacts and border effects when performing image registration and reconstruction.
The reconstructed SIO image were then computed from the acquired images using the BOSSA-SIM algorithm~\cite{lai-tim_jointly_2019}, which allows unsupervised SIM super-resolved reconstruction for thick and moving objects. It consists of minimizing the following \emph{Maximum a Posteriori} criterion, under positivity constraint:
\begin{equation}\label{eq_map_criterion}
    \begin{split}
    J(\mathbf{o}_0,\mathbf{o}_d) &=\frac{1}{2} \sum_{j=1}^N \norme {\frac{\mathbf{i}_j - \mathcal{M}_j(\mathbf{o}_0,\mathbf{o}_d)}{\boldsymbol{\sigma}_j}}_2^2 \\
    &+     \frac{\lambda}{2} \left [ \sum_{f}{\frac{|\Tilde{\mathbf{o}}_0(f)|^2}{ \mathbf{S}_{\mathbf{o}_0}(f)}} + \sum_{f}{\frac{|\Tilde{\mathbf{o}}_{d}(f)|^2}{ \mathbf{S}_{\mathbf{o}_{d}}(f)}} \right ]
    \end{split}
\end{equation}
where:
\begin{itemize}
    \item $\mathbf{o}=(\mathbf{o}_0,\mathbf{o}_d)$ is the 2-layer object to be reconstructed. It is composed of the in-focus object layer $\mathbf{o}_0$ and the defocused object layer $\mathbf{o}_d$ into which the out-of-focus contributions are rejected so as to obtain a super-resolved and optically sectioned in-focus layer $\mathbf{o}_0$;
    \item $\mathbf{i}_j$ is the SIO j-th pre-processed image;
    \item $\mathcal{M}_j(\mathbf{o}_0,\mathbf{o}_{d})
    $ is the imaging model, which will be further described below;
    \item $\boldsymbol{\sigma}_j^2$ is the noise variance, which we assume to be homogeneous here;
    \item $\lambda$ is the regularization parameter and is set to $0.3$ as explained in~\cite{lai-tim_jointly_2019};
    \item $\mathbf{S}_{\mathbf{o}_0}$ and $\mathbf{S}_{\mathbf{o}_{d}}$ are the power spectral densities of each object layer;
    \item $\Tilde{.}$ refers to the 2D discrete Fourier transform of its argument and $f$ is the 2D spatial frequency.
\end{itemize}
The noise variances $\boldsymbol{\sigma}_j^2$ and the object power spectral densities $\mathbf{S}_{\mathbf{o}_0}$ and $\mathbf{S}_{\mathbf{o}_{d}}$, are estimated from the data in an unsupervised way~\cite{Blanco-a-11,lai-tim_jointly_2019}.

The imaging model corresponding to the j-th image reads:
\begin{equation}
\begin{split}
    \mathcal{M}_j(\mathbf{o}_0,\mathbf{o}_{d})
    &= \big[\mathbf{h}_0 \star (\mathbf{m}_{j,0}. t_j[\mathbf{o}_0]) \big]_{\boldsymbol{\mathrm{III}}}(k,l) \\
    &+ \big[\mathbf{h}_d \star (\mathbf{m}_{j,d}. t_j[\mathbf{o}_d]) \big]_{\boldsymbol{\mathrm{III}}}(k,l)\label{eq_model_reduit} 
\end{split}
\end{equation}
where $\star$ depicts the discrete 2D convolution product and $t_j[.]$ is a subpixel shift operator that computes the shifted discretized object. $[.]_{\boldsymbol{\mathrm{III}}}$ is a downsampling operator that makes it explicit that the observed object $(\mathbf{o_0},\mathbf{o}_d)$ can be oversampled with respect to the SIM images $\mathbf{i}_j$ to ensure that the reconstructed super-resolved image satisfies the Shannon-Nyquist sampling theorem.
The imaging model depends of 5 parameters: the in-focus and defocused point-spread functions (PSF) $\mathbf{h}_0$ and $\mathbf{h}_d$, the in-focus and defocused illumination patterns $\mathbf{m}_{j,0}$ and $\mathbf{m}_{j,d}$ and the retinal shifts that determine the subpixel shift operator $t_j[.]$. The choice or estimation of these parameters, required for the reconstruction process, is described in Appendix~\ref{sect:choice_param}.

\section{Results}

\begin{figure*}[!htbp]
\centering
\includegraphics[width=\linewidth]{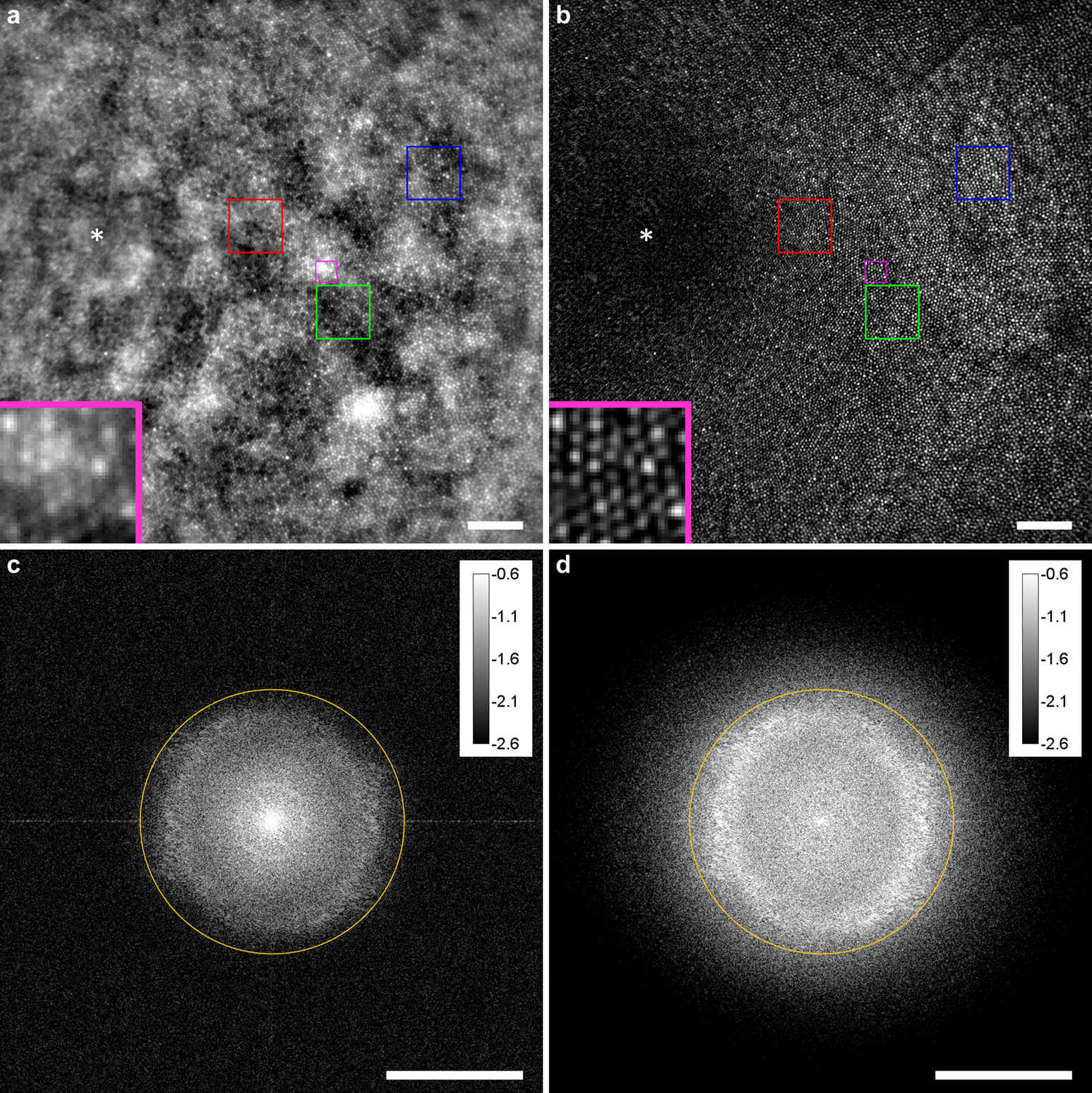}
\caption{Conventional wide-field versus SIO $2.54\degree \times2.54\degree$ retinal images, computed from the same raw data. a-d, Conventional wide-field image (a) and SIO image (b) together with their corresponding image spectra (c) and (d) respectively. Insets (a-b), magnified views of the magenta boxed $0.1\degree\times 0.1\degree$ region. Scale bars, 0.25$\degree$ (a-b), 100 cycles/degree (c-d). The displayed images are focused on the photoreceptor layer and centered at 0.9$\degree$ eccentricity. To correct for the uneven background illumination that falls off toward the edge of the field of view, a bandpass filter was applied to the wide-field image (see Appendix~\ref{sect:correction_background}). In a-b, the white asterisk marks the location of the foveal center. A linear lookup table (LUT) with a saturation of the 0.3\% of the brightest and darkest pixels is applied to a-b. In c-d, the yellow circle indicates the effective cutoff frequency of the wide-field image (98 cycles/degree), above which the corresponding spectrum is dominated by the noise. The image spectra are displayed in logarithmic scale. The blue, green and red boxed regions in a-b are magnified on Figure~\ref{fig:zoom}.}
\label{fig:fio_vs_sim}
\end{figure*}

In order to investigate the potential of SIO for super-resolved retinal imaging, we imaged the cone photoreceptors at the foveal center where their density is maximal~\cite{curcio1990human}, in a healthy subject.
79 SIO frames per orientation of the fringe patterns were processed to obtain the reconstructed SIO image. SIO raw frames are shown in Figure~S1. 
To compare SIO with conventional wide-field ophthalmoscopy under the same imaging conditions (light flux, AO correction, eye movement), a wide-field image was constructed by averaging the registered SIO raw images. We checked that the illumination patterns are effectively removed by this averaging and registration process so as to produce a reliable conventional wide-field image. The comparison between the wide-field and SIO images is shown in Figure~\ref{fig:fio_vs_sim}.
In the wide-field image (Figure~\ref{fig:fio_vs_sim}a), even after correction of the uneven background illumination (Figure~S2), the contrast of the cone mosaic is dominated by the out-of-focus background, which comes from defocused layers and choroidal scattering~\cite{burns_adaptive_2019}.
In the SIO image (Figure~\ref{fig:fio_vs_sim}b), this defocused contribution is rejected and the cones are more visible. Thus, SIO achieves optical sectioning, which greatly reduces the scattering light background and consequently improves the contrast of the in-focus retinal layer.

To evaluate the resolution of the wide-field and SIO images, we computed their image spectra (Figure~\ref{fig:fio_vs_sim}c-d).
In the wide-field image spectrum, the diffraction-limited cutoff frequency 
is not reached because of the noise. We thus defined the wide-field effective cutoff frequency as the radial frequency above which the object's contribution to the image power spectral density becomes smaller than the noise level.
We have evaluated the effective cutoff spatial frequency of the conventional wide-field image $f_{c,\text{eff}}^{WF}$ from the noise and object power spectral densities (PSD), which were estimated using the unsupervised method proposed in BOSSA-SIM~\cite{lai-tim_jointly_2019}.
The measurements of this effective cutoff frequency yields $f_{c,\text{eff}}^{WF}=98$ cycles/degree.
This value is symbolized by the yellow circle in Figure~\ref{fig:fio_vs_sim}c-d.
In the SIO image, the frequency bandwidth is extended by about the value of the pattern frequency $f_m=34$~cycles/degree beyond the wide-field frequency bandwidth, thus leading to a SIO effective cutoff frequency of $f_{c,\text{eff}}^{SIO}=132$~cycles/degree.
\noteylt{Hence, SIO achieves super-resolution with respect to the effective resolution of the conventional wide-field image.}

\begin{figure*}[!htbp]
\centering
\includegraphics[width=0.95\linewidth]{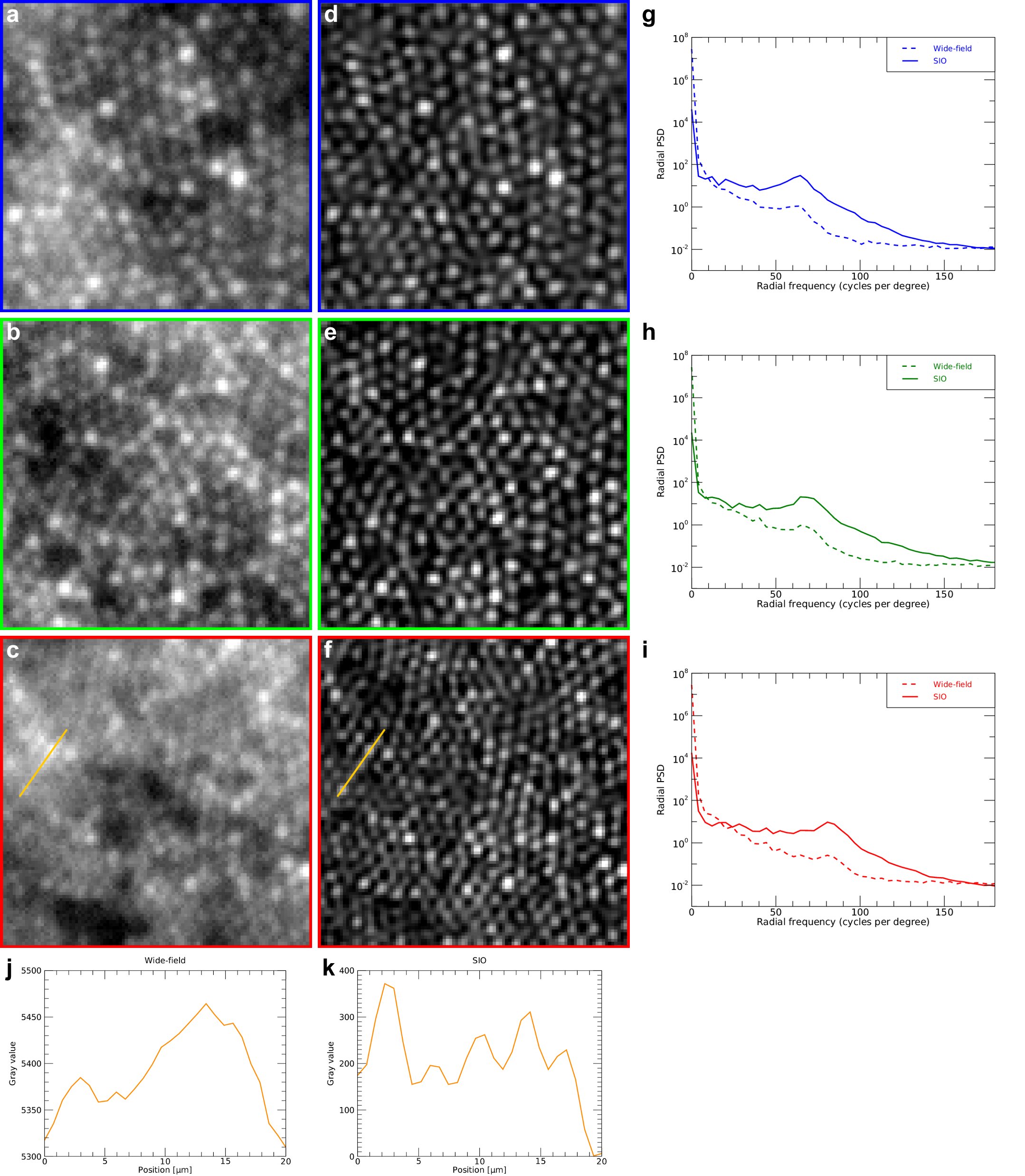}
\caption{Comparisons of the wide-field and SIO retinal images over $0.25\degree \times0.25\degree$ enlarged views. a-c, Magnification of the blue (a), green (b) and red (c) boxed regions in the wide-field image (Figure~\ref{fig:fio_vs_sim}a), d-f, Corresponding magnification from the SIO image (Figure~\ref{fig:fio_vs_sim}b), g-i, Radial power spectra of the blue (g), green (h), respectively red (i) boxed regions, j-k, Plots of the image intensity along the 20 $\mu$m lines drawn in c (j) and f (k). The blue, green and red boxed regions are centered at retinal eccentricities of 1.6\degree{}, 1.2\degree{}, and 0.7\degree{} respectively. A linear LUT with a saturation of the 0.3\% of the brightest and darkest pixels is applied to the images displayed in a-f.}
\label{fig:zoom}
\end{figure*}
Figure~\ref{fig:zoom} provides further comparisons between the wide-field and the SIO images over enlarged views of $0.25\degree \times0.25\degree$ field-of-view.
The radial power spectra plotted in Figure~\ref{fig:zoom}g-i, exhibits a peak at 65 cycles/degree (g), 69 cycles/degree (h) and 81 cycles/degree (i) related to the photoreceptor average density in each of the $0.25\degree \times0.25\degree$ regions of interest. These peaks reach higher values for the SIO, which shows that the cone mosaic is higher contrast in the SIO image.
At $0.7$\degree eccentricity (Figure~\ref{fig:zoom}c and f), the \ac{sio} is able to resolve cones that are indistinguishable in the wide-field image, as exemplified in the intensity plots (Figure~\ref{fig:zoom}j-k). In particular, the spacing of the last two cones resolved only in the SIO intensity plot corresponds to a spatial frequency of about $110$ cycles/degree, which is higher than the wide-field effective cutoff frequency. 


\section{Discussion}

The higher resolution and contrast achieved by SIO compared with conventional wide-field ophthalmoscopy makes SIO a highly valuable tool for imaging foveal cones and monitoring cone density, which is a key biomarker for detecting the early onset of common retinal degenerative diseases. To investigate this matter, we compared the cone density map computed from the SIO reconstructed image with the one computed from the conventional wide-field image, as shown in Figure~\ref{fig:density}. The computation of the cone density maps is detailed in Appendix~\ref{sect:cone_map}.

\begin{figure*}[!hbt]
    \centering
    \includegraphics[width=\linewidth]{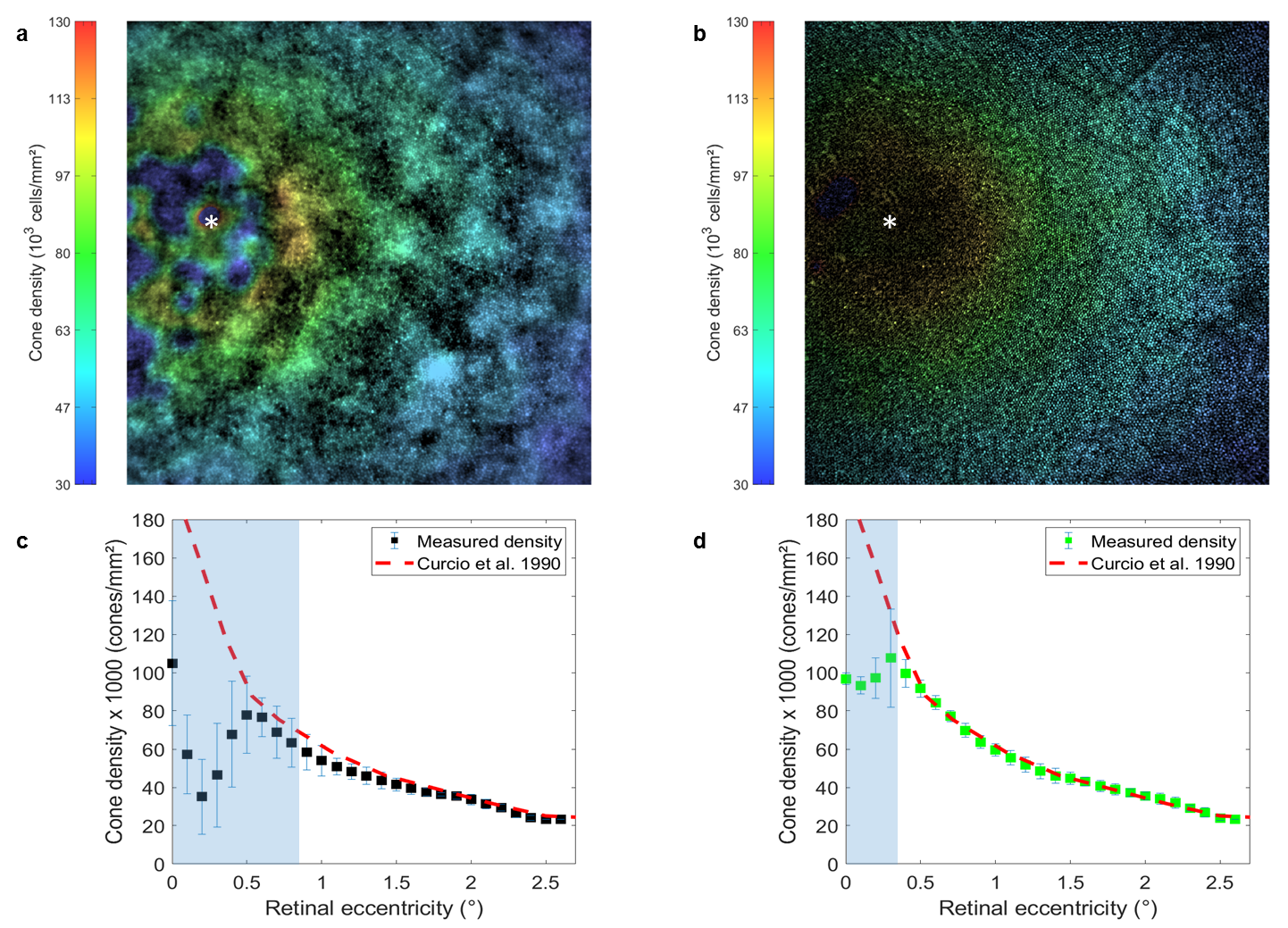}
    \caption{Measurement of the cone density. a-b, Cone density distribution color maps obtained from the wide-field image (a) and from the SIO image (b). c-d, The estimated cone density mean (squares) and standard deviation (blue lines) values as a function of the retinal eccentricity respectively computed from a and b. The white asterisk in a,b, indicates the foveal center. The red dashed line corresponds to histological measurement from an average retina~\cite{curcio1990human}. The blue shaded region in c,d, indicates the eccentricities for which the cone density estimated from each image is unreliable (ie., when the ratio of standard deviation over the mean value is greater than $0.1$).}
    \label{fig:density}
\end{figure*}

For both images, the measurements are consistent with histology up to a retinal eccentricity below which the estimated densities start to deviate as the cones are no longer resolved. We defined the retinal eccentricity below which the measurements are unreliable as that for which the ratio of the standard deviation over the mean value of the estimated cone density becomes greater than 10\% (blue shaded region in Fig.~\ref{fig:density}(c-d)).
Using this criterion, it clearly appears that the SIO image allows a much more accurate cone density measurement than the conventional wide-field image, increasing the range of eccentricities for which the estimated densities are reliable.
Moreover, unlike scanning ophthalmoscopes, SIO images are distortion-free, which is of the utmost importance for longitudinal studies of cone density.

\begin{figure*}[!htbp]
\centering
\includegraphics[width=0.9\linewidth]{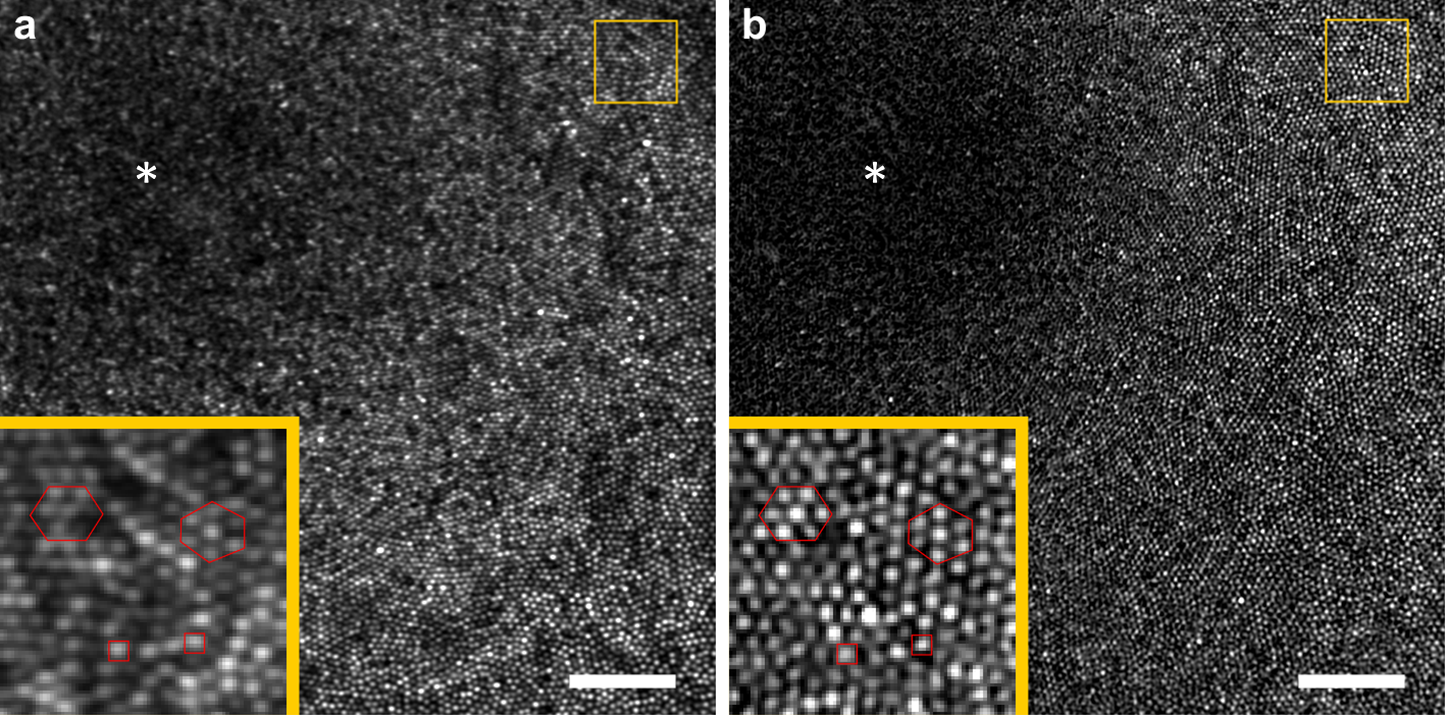}
\caption{cSLO versus SIO $1.74\degree \times1.74\degree$ retinal images. a, cSLO image, b, SIO image. Scale bars, 0.25\degree. The insets display a magnification of the yellow boxed $0.2\degree\times 0.2\degree$ regions. In boths insets, the red hexagone and square shapes highlighting common cones were drawn to facilitate the image comparison. The white asterisk indicates the foveal center. A linear LUT with a saturation of the 0.3\% of the brightest and darkest pixels was applied to both images.}
\label{fig:aoslovssim}
\end{figure*}

Figure~\ref{fig:aoslovssim} shows how the SIO image compares favourably with a typical AO-corrected \ac{cslo} image. Indeed, the cones appear more contrasted and sharper in the SIO image. It should be noted that the cone brightness is not the same from one imaging modality to the other due to the temporal variability of photoreceptor reflectance~\cite{pallikaris_reflectance_2003,cooper_spatial_2011}.
Additionally, SIO and \ac{cslo} qualitatively achieve a similar optical sectioning. This result was expected, as it was already demonstrated in microscopy that \ac{sim} and confocal imaging enable similar axial sectioning~\cite{wilson_optical_2011}.
More information about the cSLO that we used are provided in Appendix~\ref{sect:aoslo}.

\section{Conclusion}

In conclusion, we have developed SIO, an adaptive optics-corrected wide-field ophthalmoscope using structured illumination that achieves super-resolution and optical sectioning in the living retina, without distortion over a large field-of-view. The SIO exploits the BOSSA-SIM reconstruction algorithm, which presents five main features that make it successful in reconstructing high resolution contrasted retinal images.
Firstly, it is based on an imaging model that takes into account the retinal shifts.
Secondly, its original 2-layer object model, which distinguishes the in-focus signal from the out-of-focus contribution, enables one to jointly achieve optical sectioning and super-resolution from two-dimensional data.
Thirdly, thanks to the Bayesian framework, the method is robust to noise, which can be important in wide-field retinal images. Fourthly, its reconstruction hyper-parameters (object and noise power spectral densities) are adjusted in an unsupervised fashion (\textit{i.e.}, automatically) from the data.
Lastly, the method imposes a positivity constraint on the reconstructed object, which is known to induce spectral extrapolation for objects on a dark background. Thus, this positivity constraint, combined with the optical sectioning, contributes to an extension of the reconstruction frequency bandwidth (additional to the super-resolution brought by \ac{sim}).
By imaging the cone mosaic near the fovea, we have shown contrast and resolution enhancement in the SIO reconstructed image compared with a conventional wide-field image.
\noteylt{It should be noted that the resolution improvement is currently limited by the poor \ac{snr} of the raw images, which do not reach the diffraction limit due to the noise.
Ongoing developments are aiming at improving the \ac{snr} of the raw images in order to further increase the resolution enhancement enabled by \ac{sio}.}

Combining super-resolution and distortion-free acquisition is highly desirable for imaging a neuro-vascular network such as the retina, which is why it has been attempted several times in the past \cite{gruppetta_theoretical_2011, shroff_structured_2010}.
We expect this breakthrough to change the balance between wide-field and scanning systems in retinal imaging, where scanning systems are more common. Indeed, we have used a wide-field ophthalmoscope and demonstrated super-resolution in the eye with neither photon loss, which occurs in small pinhole confocal scanning systems, nor distortion, which is inherent to any scanning system. This paves the way towards accurate photoreceptor mapping in challenging patients with strong retinal motion or with a small pupil.
More generally, it provides a generic strategy to achieve super-resolution and optical sectioning in a wide variety of biomedical imaging applications, despite sample motion, scattering and aberrations.

\section*{Appendix}
\appendix


\section{Choice of the parameters for the SIO reconstruction}\label{sect:choice_param}

The parameters' choice of the imaging model (Equation~(2)), required for the SIO reconstruction, is described below.

The PSFs are derived from the corresponding optical transfer functions (OTF) in the reciprocal space. The in-focus OTF $\Tilde{\mathbf{h}}_0$ is modeled by a diffraction-limited OTF attenuated by an exponential term with a dampening parameter of $0.1$ as in FairSIM~\cite{muller_open-source_2016} to account for residual optical aberrations. The out-of-focus OTF $\Tilde{\mathbf{h}}_d$ is obtained by applying the same exponential attenuation to an OTF with defocus aberration. The defocus value was chosen in such a way that the defocused OTF value at the modulation spatial frequency $f_{m}$ is null.

As we project only two sinusoidal patterns, one oriented at $\theta_1=-45\degree$ and the other at $\theta_2=+45\degree$, there are two sets of patterns $\{\mathbf{m}_{\theta,0},\mathbf{m}_{\theta,d}\}_{\theta}$ to be estimated (See Equation~(2)). The in-focus illumination pattern at a given orientation $\theta$, is given by:
\begin{equation}\label{eq:m_illu_o0}
    \mathbf{m}_{\theta,0}(\vec{r})= \mathbf{I}_{\theta}(\vec{r}) [1 + \boldsymbol{\alpha}_{\theta}(\vec{r}) \cos(2\pi \vec{f}_{\theta,m}.\vec{r}+\phi_{\theta})]
\end{equation}
with $\vec{r}=(k,l)$ the 2D spatial coordinates.
The intensity map $\mathbf{I}_{\theta}$ takes into account the heterogeneous background due to inhomogeneities in the illumination beam and to multiple scattering effects inside the eye (see Figure~\ref{fig:aofio_filtering}). $\mathbf{I}_{\theta}$ was estimated by fitting a third-order polynomial on the average SIO frame for the given $\theta$.
To account for variable modulation contrasts over the field of view, a contrast map $\boldsymbol{\alpha}_{\theta}$ of values between $0.1$ and $0.3$ was set.
Finally, the spatial frequency $\vec{f}_{\theta,m}$ and the phase $\phi_{\theta}$ of the illumination pattern were estimated using a least square fit of a sinusoidal fringe model over a filtered version of the average SIO frame for the given $\theta$. The purpose of the filtering is to filter out the object structures that may skew the pattern parameter fitting.

As we incoherently project a fringe pattern onto the retina, the fringe contrast decreases with defocus and thus, the illumination patterns are only contrasted at the in-focus object plane. Hence, we can consider that the defocused illumination pattern $\mathbf{m}_{\theta,d}$ is not modulated: 
\begin{equation}\label{eq:m_illu_od}
    \mathbf{m}_{\theta,d}(\vec{r})= \mathbf{I}_{\theta}(\vec{r})
\end{equation}

The retinal shifts were estimated in two steps: the SIO raw images were first filtered to cut off the spatial frequencies of the illumination patterns so as to obtain approximate wide-field images (without the fringe pattern), then used as inputs in a subpixel shift estimation method~\cite{blanco_registration_2014}. 

\section{Correction of the heterogeneous background in the AOFIO image}\label{sect:correction_background}

\begin{figure*}[!htbp]
    \centering
    \includegraphics[width=\linewidth]{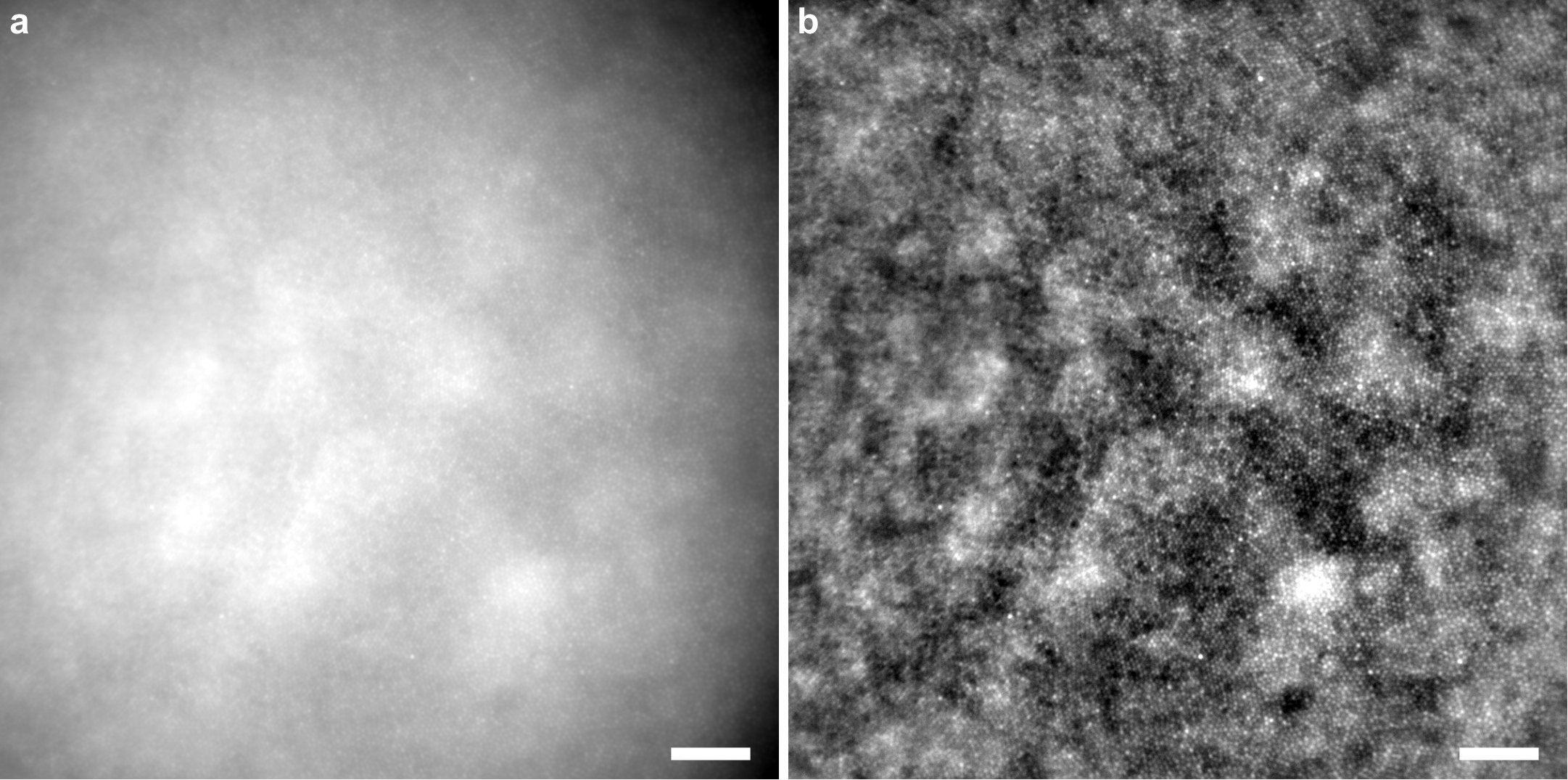}
    \caption{Correction of the heterogeneous background in the conventional wide-field image. a, $2.54\degree \times2.54\degree$ wide-field image, b, Wide-field image after background illumination correction. Scale bars, 0.25\degree. The corrected image was obtained by filtering the wide-field image (a) using ImageJ's "bandpass filter" function. The parameters were set to filter large structures down to 100 pixels and small structures up to 0 pixels. A linear LUT with a saturation of the 0.3\% of the brightest and darkest pixels was applied to both images.}
    \label{fig:aofio_filtering}
\end{figure*}

Due to the illumination inhomogeneities and mostly to the scattering of the light when propagating inside the eye, retinal wide-field images contain a strong background which slightly decreases toward the edges of the field. As shown in Figure~\ref{fig:aofio_filtering}(a), this background significantly reduces the contrast of the retinal structures that are imaged (in our case, the foveal cones). As the background is composed of low spatial frequencies, we can filter it out by applying a highpass filter to the wide-field image. The resulting filtered image is displayed in Figure~\ref{fig:aofio_filtering}(b).
One drawback of this correction is that it also suppresses the low spatial frequency of the observed retinal layer.

\section{Computation of the cone density map}\label{sect:cone_map}

To generate a pointwise density map, we divided the cone mosaic image into an overlapping grid of $64\times 64$ pixel (corresponding to $88~\mu m \times 88~\mu m$) regions of interest (ROIs), where each ROI was displaced from the previous one by $10$ pixels (corresponding to $15 \mu m$). These values were chosen empirically to provide a good trade-off between pointwise accuracy and map smoothness. Then, cone density was computed for each ROI using a fully automated algorithm based on modal spacing as described in \cite{cooper_fully_2019}. Bicubic image interpolation was then used to increase the size of the cone density map in order to match the cone mosaic image.

\section{Acquisition of the cSLO retinal images}\label{sect:aoslo}

In order to compare SIO with a confocal scanning laser ophthalmoscope, additional retinal images were acquired with a commercial AO-corrected cSLO system (MAORI, Physical Sciences Inc., Andover MA USA)~\cite{grieve_vivo_2018}. Image acquisition was performed on the same subject as for the SIO and in the same conditions, \textit{i.e.}, with neither pupil dilation nor cyclopegia, in a dark room. 
The optical power entering the eye was set to $800 \mu W$ at $780$ nm for the illumination source and $700 \mu W$ for the wavefront sensor source at $840$ nm.
100 cSLO $2\degree \times 2\degree$ individual frames centered on the fovea were acquired at 24 Hz.
They were then processed using the MAORI image processing library for registration and distortion compensation~\cite{mujat_highresolution_2015}. Figure~6(a) of the primary document shows the processed cSLO image, which was manually registered with the SIO image.



\section*{Funding}
This work was supported by the HELMHOLTZ synergy grant (European Research Council (ERC) [610110]) and the French nation fund LIGHT4DEAF [ANR-15-RHUS-0001].

\section*{Acknowledgments}
The authors would like to thank José-Alain Sahel for clinical expertise and support, and Frédéric Cassaing for fruitful discussions.

\section*{Disclosures}
The authors declare that there are no conflicts of interest related to this article.





\bibliography{bib_simletter}


\end{document}